\input amstex
\documentstyle{amsppt}
\topmatter
\title{On several correlation integrals of the deep level transients}\endtitle
\author{Pham Quoc Trieu and Hoang Nam Nhat}\endauthor
\date{Faculty of Physics, Hanoi University of Sciences, VNU Hanoi}\enddate
\curraddr{}\endcurraddr
\email{namnhat@hn.vnn.vn}\endemail
\thanks{The authors would like to thank the Center for Materials Science, HUS - VNU HN for supports  regarding the work on the DLTS and other equipments }\endthanks

\abstract{This works presents the theoretical study on several correlation integrals of the capacitance transients of the deep levels. The deconvolution of the transient signals was the major subject of the number of methods referred to under the common name as the Deep Level Transient Spectroscopy methods. In general the separation of the overlapping exponential decays $C(t)$ does not provide a unique solution, so the detection of the closely spaced deep levels by these transients should base mainly on the temperature dependence $C(T)$, not only on the $C(t)$. The results show that the average emission factor $e_n$ is obtainable directly from various correlation integrals of the capacitance transients and the average activation energy $E$ of the deep levels is detectable via the shift operators of the transients according to the temperature. A new scanning technique is suggested.}\endabstract
\endtopmatter

\head{}Nomenclature\endhead
We call a {\it normalized capacitance} $C_n(t)$ at fixed temperature $T$ the function $C_n(t)=C_0^{-1} \times [C(t)-C_1]$, where $C_0$ is $C(t)$ at $t=0$ and $C_1$ is $C(t)$ at $t=\infty$. For $0<t<\infty$, $C_n(t)$ always specifies the relation $0<C_n(t)<1$ , i.e. $L(t) = Ln[C_n(t)]$ has definite and negative value within (0,1). Taking $Ln$ on $L(t)$ is not possible but $M(t) = Ln[-Ln[C_n(t)]]$ has definite values. By replacing $t$ by $T$ we have a normalized capacitance $C_n(T)$ at fixed gate $t$, this $C_n(T)$ is not exponential. For section 3, $w=E/k$ where $E$ is the activation energy of the deep level and $k$ is the Boltzman constant.

\head {1. Introduction} \endhead
The problem of separation of the closely spaced levels in Deep Level Transient Spectroscopy has last for decades but no clear answer is available until now. Several people suspected that the problem is ill-posed in principle [4]. By its nature the problem finalizes in the search for a unique decomposition of any exponential decay into a finite combination of several other ones. Theoretically all exponential functions is  expandable into an infinite series of the other exponential components, so we may expect the existence of many finite approximations which satisfy the precision limit. Thus the critical question follows: {\it if there exist many decompositions so which of them preserves the physical meaning, or is that the true physical reality what the DLTS discovers?}

The DLTS in its nature is not a purely measurement technique but involves a lot of analysis, so the obtained results can hardly be attached directly to the physical reality without some theoretical presumptions. One of the basic presumptions was about the duality of the DLTS peaks and the emission factors of the deep centers. This was so evident that needed no explicit statement, however we must re-state here that the word {\it "emission center"} must be understood statistically, i.e. the DLTS peak reflects only the {\it statistical behaviour} of the underlying physical reality and not the physical reality (of the emission centers) themselves. Thus giving a set of emission centers, the appearance of DLTS peaks follows the statistical rules for composition of the output signals which must not be a simple one-to-one relation in principle. 

There are known various interpretations of the  level resolution [2]. Providing that the deep levels are statistical terms, the level closeness must also be understood statistically, i.e. closeness as the smallest observable variation in the statistical behaviours of the levels. We may call this closeness the {\it local closeness}. The physical meaning of this definition is that in some temperature range the statistical contributions of the levels to the output signals are not comparable, whereas in some other range they are quite undistinguishable. Some levels may be seen as the closely spaced within one $T$ range but not out of it. Consequently, if only one level contributes to the output signal while all others remain inactive then this level is far from the rest and is separable by the sense of the underlying DLTS technique. As the evident consequence, the local closeness will require more smoother temperature scan than for the normal techniques (e.g. for D.V. Lang's classical DLTS [6] or for S. Weiss and R. Kassing's Fourier DLTS [7]).

Now let us turn to the problem of the statistical rules for composition of the final output signals. Until now no one hesitates to write:

$C_n$(t)= $\sum\limits^n_{i=1}e^{-te_{ni}}$   (1.1)

which automatically assumes the composition additiveness. Although this relation has provided satisfied results when the levels are well separated, there is no clear reason for the additiveness in case of the closely spaced levels. As known, for the perfect exponential decays, the level contributions are not statistically independent (the known Levy theorem states that for any two Gaussian distributions, and thus the exponential ones, the final composition is always Gaussian, see [5]). If one assumes the final emission factor to be $e_{final}$ = $\sum\limits^n_{i=1}p_i(e)e_{ni}$  where $p_i(e)$ is the statistical weight (emission factor's density distribution) then final output $C_n(t)$ evidently becomes:

$C_n(t)= e^{-te_{final}} = \prod\limits^n_{i=1}e^{-tp_i(e)e_{ni}}$ (1.2)

It is questionable whether (1.1) or (1.2) takes place. With regards to the above stated problems, we present in this article the study on several correlation integrals of the output signals $C(t)$ and $C(T)$ and suggest a scan technique which bases mainly on scanning the $C(T)$ according to the temperature. As shown below, the emission factor and activation energy may be retrieved directly from the statistical study of the output. Evidently these values correspond to the locally closed states if these states exist.

\head {2. Correlation integrals} \endhead
By definition all correlation functions $R(\tau)$ of a signal $F(t)$ involve the shift operator of that signal ${\hat \bold T}_{\tau}[F(t)]=F(t-{\tau})$ and represent the average of signal over the preset period ${\tau}$. Naturally, the correlation functions filter noise, but also remove the fine local structure of the signals (see e.g. [3]).

\subhead{2.1. Correlation integrals at fixed T}\endsubhead

Taking correlations at fixed $T$ defacto means averaging signal according to time $t$. At fixed temperature the development of signal after a time $t$ wholly depends on the emission constant $e_n$ and has a simple exponential form $e^{-e_nt}$. Let ${\lambda}$ be a period width and integrate through the whole period.
\subsubhead{2.1.1. Cross-correlation of $C_n(t)$ and $\frac{1}{C_n(t)}$}\endsubsubhead

$R({\tau})=\frac {1}{\lambda}\int\limits^{\lambda/2}_{-\lambda/2} \frac{e^{-e_nt}}{e^{-e_n(t-\tau)}}dt=e^{-e_n{\tau}}\Rightarrow -\frac{1}{\tau}Ln[R({\tau})]=e_n$ 

(for all $T$)

\subsubhead{2.1.2. Autocorrelation of $L(t)$}\endsubsubhead

$R({\tau})=\frac{1}{\lambda}\int\limits^{\lambda/2}_{-\lambda/2}(-e_nt)(-e_nt+e_n{\tau})dt=\frac{(e_n{\lambda})^2}{12}$ 

\subsubhead{2.1.3. Autocorrelation of  $-\frac{L(t)}{t}$}\endsubsubhead

$R({\tau})=\frac{1}{\lambda}\int\limits^{\lambda/2}_{-\lambda/2}(\frac{-e_nt}{t}) (\frac{-e_nt+e_n \tau}{t-\tau}) dt=e^2_n$ 

One may also check that a cross-correlation of $L(t)^{-\frac{1}{t}}$ and $ \frac{1}{L(t)^{-\frac{1}{t}}}$ is always 1.

$R({\tau})=\frac{1}{\lambda}\int\limits^{\lambda/2}_{-\lambda/2} (\frac{-e_nt}{t}) (\frac{t-\tau}{-e_nt+e_n \tau}) dt=1$ 

\subhead{2.2. Correlation integrals at fixed $t$}\endsubhead

This class of correlation functions illustrates the averaging process according to temperature $T$, i.e. the process of filtering the temperature noise. Write $C_n(T) = e^{-t  {\rho}  T^2 e^{-w/T}}$,  $L^{-1/t}(T) = {\rho} T^2 e^{-w/T}$ and put $M(T)=Ln[-L(T)^{-1/t}]=Ln({\rho})+2Ln(T)-{\omega}/T$. The correlation is considered within one segment of $T$ so we integrate from $T_1$ to $T_2$. Let $\Delta T=T_2-T_1$.

The cross-correlation between  $M(T)$ and $1/M(T)$ is:

$R({\tau})=\frac{1}{\Delta T}\int\limits^{T_2}_{T_1} {\rho} T^2e^{-\frac{\omega}{T}} {\rho} (T-{\tau})^2e^{-\frac{\omega}{(T-{\tau})} }dT=$ 

$=\frac{1}{\Delta T}\int\limits^{T_2}_{T_1}[\frac{T}{T-{\tau}}]^2e^{-(\frac{\omega}{T}-\frac{\omega}{T-{\tau}) } }dT$

Substitute $x=\frac{1}{T}, x_1=\frac{1}{T_1}, x_2=\frac{1}{T_2}$ and $\Delta x = x_2-x_1$. In unit of $x$ and replace $\sigma$ for ${\tau}$ as a new shift:

$C(x)=e^{-t {\rho} x^{-2}e^{-{\omega} x}}$, $L(x)=-t {\rho} x^{-2}e^{-{\omega} x}$, $L(x)^{-\frac{1}{t} }={\rho} x^{-2}e^{-{\omega} x}$ and the correlation integral is:

$R({\sigma})=\frac{1}{\Delta x}\int\limits^{x_2}_{x_1} \frac{{\rho} x^{-2}e^{-{\omega} x}}{ ({\rho} (x-{\sigma})^{-2}e^{-{\omega} (x-{\sigma})} } dx=\frac{e^{-{\omega}{\sigma}}} {\Delta x} \int\limits^{x_2}_{x_1}(1-\frac{\sigma}{x})^2dx$

\

After integration:

$R({\sigma})=e^{-{\omega}{\sigma}}[1+2{\sigma}\frac{Ln(x_1/x_2)}{\Delta x}+ {\sigma}^2(\frac{1}{x_1x_2})]$

\

By putting $A=\frac{1}{\Delta x}Ln(\frac{x_1}{x_2})$, $B=\frac{1}{x_1x_2}$ we obtain:

\

${\omega} = \frac{1}{\sigma}Ln[\frac{1+2{\sigma}A +{\sigma}^2 B}{R({\sigma})}]$

\

Therefore the correlation integral directly determines the activation energy of the deep level.

\head{3. Shift operator $C_n(T \times p) = \hat \bold T_p[C_n(T)]$}\endhead

According to temperature $T$ the shift operator $ \hat \bold T_p$ moves $C_n(T)$ onto $C_n(T \times p)$ for a real positive multiplicative constant $p: C_n(T \times p) = ${\bf \^{T}}$_p[C_n(T)]$. In the following sections {\bf \^{T}}$_p$ will be derived (many aspects of the shift operators are explored in [1]).

\subhead{3.1. Shift operator of $Cn(T)$, $Ln(T)$ and $Mn(T)$}\endsubhead

Dividing $C_n(T {\times} p)=e^{-t {\rho} (T{\times}p)^2e^{-\frac{\omega}{T {\times} p}} }$ by $C_n(T)=e^{[-t {\rho}T^2e^{-\frac{\omega}{T}} ]}$ leads to:

$C_n(T {\times} p) = {\hat \bold T}_p[C_n(T)] = [C_n(T)]^{p^2e^{-\frac{\omega}{T}(\frac{1}{p}-1)} }$

Similarly:

$L(T {\times} p) = {\hat \bold T}_p[L(T)] = [L(T)]p^2e^{-\frac{\omega}{T}(\frac{1}{p}-1)}$   (3.1.1)

$M(T {\times} p) ={\hat \bold T}_p[M(T)] = M(T)+2Ln(p) - \frac{\omega}{T}(\frac{1}{p}-1)$

\subhead{3.2. Calculation of energy using shift operators}\endsubhead

With ${\omega}=E/k$ and by relation (3.1.1):

${\omega}=T(\frac{p}{1-p})Ln[p^2 \frac{L(T)}{ {\hat \bold T}_p[L(T)]} ]$     (3.2.1)

So an arbitrary shift from $L(T)$ to $L(T \times p)$ determines a energy constant significant within this temperature shift [$T \rightarrow T{\times}p$]. A technique for detection and analysis of closely spaced levels by scanning ${\omega}$ in various temperature range is suggested. We call it the 'Selective Temperature Scan Technique' (STST).

\head{4. Conclusion}\endhead

Not only ${\omega}$ but also ${\rho}$ is detectable via shift operators. For calculation purposes we have simplified (3.2.1) to the following relations. Suppose a shift from $(T_i,t_i)$ to $(T_k,t_k)$ and use short notes $L_i=L(T_i), L_k=L(T_k)$.

${\omega}_{ik}=(\frac{T_iT_k}{T_k-T_i})Ln[(\frac{L_k}{L_i}) (\frac{t_i}{t_k}) (\frac{T_i}{T_k})^2]$     (2.2.1)

${\rho}_{ik}=[(\frac{t_k^{T_k}}{t_i^{T_i}}) (\frac{T_k^{T_k}} {T_i^{T_i}})^2 (\frac{(-L_i)^{T_i}} {(-L_k)^{T_k}})]^{\frac{1}{T_i-T_k}} $

In practical case we put $t_i=t_k$ and chose $T_k$ as close as possible to $T_i$ , i.e. $T_k$ displaces from $T_i$ only by one scan step. The displacement $T_k-T_i$ refers to the sensibility limit of the temperature setting and the scanning range $[T_1...T_2]$ refers to the statistical weight. For general case where $t_i{\not=}t_k$, the $t_k-t_i$ means the time variation of the statistical weight. The functionality of method is adjustable by these three parameters. Note that the above relations hold only for cases where the statistical contribution of one deep center (in the scanning temperature range) is significantly greater than of all the rest.

\head{References}\endhead

{\noindent [1] Cartier P., Mathemagics. In: M. Planat, editor. Lecture Notes in Physics 550, Springer-Verlag, 2000, p. 6-67}

{\noindent [2] Doolittle W.A., Rohatgi A. A new figure of merit and methodology for quantitatively determining defect resolution capabilities in deep level transient spectroscopy analysis. J. Appl. Phys. 1994; 75(9):4570-4575.}

{\noindent [3] M. Schwartz. Comm. Systems And Techniques. McGraw-Hill NY, 1966.}

{\noindent [4] Dobaczewski L., Peaker A.R. Laplace DLTS - an overview. 1997. Internet resource: 
http://www.mcc.ac.uk/cem/laplace/laplace.html}

{\noindent [5] Lévy P. Calcul des probabilités. Gauthier Villars Paris, 1930.}

{\noindent [6] Lang D.V.  Deep level transient spectroscopy: A new method to characterize traps in semiconductors. J. Appl. Phys. 1974; 45: 3023-3032.}

{\noindent [7] Weiss S., Kassing R. Deep level transient fourier spectroscopy (DLTFS) - a technique for the analysis of deep level properties. Solid-St Electron 1988; 31(12): 1733-1742.}

\end{document}